\documentclass[10pt,aps,prb,showpacs,floatfix,raggedbottom,floats,superscriptaddress,twocolumn]{revtex4-1}
\usepackage{graphicx,latexsym}
\usepackage{dcolumn}
\usepackage{amssymb,amsmath,bm}
\usepackage{color}
\usepackage[normalem]{ulem}

\DeclareMathOperator{\vctrz}{vec}

\begin{document}

%-----------------------------------------------------------------
\title{Time-dependent current into and through multilevel parallel quantum dots\\ 
       in a photon cavity}

\author{Vidar Gudmundsson}
\email{vidar@hi.is}
\affiliation{Science Institute, University of Iceland, Dunhaga 3, IS-107 Reykjavik, Iceland}
\author{Nzar Rauf Abdullah}
\affiliation{Physics Department, College of Science, 
             University of Sulaimani, Kurdistan Region, Iraq}
\author{Anna Sitek}
\affiliation{Science Institute, University of Iceland, Dunhaga 3, IS-107 Reykjavik, Iceland}
\affiliation{Department of Theoretical Physics, Wroc{\l}aw University of Science and Technology, 50-370 Wroc{\l}aw, Poland}
\author{Hsi-Sheng Goan}
\email{goan@phys.ntu.edu.tw}
\affiliation{Department of Physics and Center for Theoretical Sciences, National Taiwan University, 
             Taipei 10617, Taiwan}
\affiliation{Center for Quantum Science and Engineering, 
             National Taiwan University, Taipei 10617, Taiwan}
\author{Chi-Shung Tang}
\email{cstang@nuu.edu.tw}
\affiliation{Department of Mechanical Engineering, National United University, Miaoli 36003, Taiwan}
\author{Andrei Manolescu}
\email{manoles@ru.is}
\affiliation{School of Science and Engineering, Reykjavik University, Menntavegur 
             1, IS-101 Reykjavik, Iceland}

%
%----------------------------------------------------------------

\begin{abstract}
We analyze theoretically the charging current into, and the transport current through, a nanoscale two-dimensional 
electron system with two parallel quantum dots embedded in a short wire placed in a photon cavity. A plunger gate 
is used to place specific many-body states of the interacting system in the bias window defined by the external leads. 
We show how the transport phenomena active in the many-level complex central system strongly depend on the gate voltage.
We identify a resonant transport through the central system as the two spin components of the one-electron ground
state are in the bias window. This resonant transport through the lowest energy electron states seems to a large
extent independent of the detuned photon field when judged from the transport current. This could be expected in the 
small bias regime, but an observation of the occupancy of the states of the system reveals that this picture is
not entirely true. The current does not reflect slower photon-active internal transitions bringing the system into
the steady state. The number of initially present photons determines when the system reaches the real steady state. 
With two-electron states in the bias window we observe a more complex situation with intermediate radiative and 
nonradiative relaxation channels leading to a steady state with a weak nonresonant current caused by inelastic 
tunneling through the two-electron ground state of the system. The presence of the radiative channels makes this 
phenomena dependent on the number of photons initially in the cavity.  
\end{abstract}

\maketitle
%
%----------------------------------------------------------------------------------------
%
Various properties of nanoscale electron and spin systems in microwave cavities
are presently the focus point of many researchers. Just to mention some; photon emission
from a cavity-coupled double quantum dot caused by an electron transport through it
has been reported,\cite{PhysRevLett.113.036801} and the manipulation of spin 
qubits in cavities has gained paramount interest.\cite{Peterson380:2012,Science.349.2015}

Investigations of transport of electrons through solid-state electronic systems
placed in photon cavities are gaining attention. Partially, this is due to the obvious
connection to efforts to achieve quantum computation in a solid-state system, 
and partially it is due to the interest to study fundamental light-matter interactions
in a system expected to be highly tunable and offer increased sensitivity 
of measurements. In several cases the electronic systems have been single or
multiple quantum dots created with InAs,\cite{Peterson380:2012}
GaAs,\cite{Frey11:01,PhysRevLett.110.066802} carbon nanotubes,\cite{Science.349.2015} 
or graphene,\cite{PhysRevLett.115.126804}
and very recently in SiGe heterostructures.\cite{Mi}

Experiments have been reported on
carbon nanotube quantum dots in a planar microwave cavity coupled to external fermionic or 
superconducting leads. The sensitivity of the measurements due to the cavity 
allows for the detection of a photon assisted current of 0.3 pA corresponding 
to the mean photon number in the cavity being 120,\cite{Delbecq11:01,PhysRevX.6.021014} 
a current that is much lower than what is common in measurements of photon assisted tunneling through 
quantum dots when they are not placed in a cavity.\cite{PhysRevLett.73.3443}

Numerous models have been presented for transport and processes in cavity-quantum electrodynamics
systems,\cite{PhysRevB.81.155303,PhysRevB.87.195427,PhysRevB.90.085416} and 
time-dependent electron transport through nano-electron systems in a photon cavity 
in the transient,\cite{doi:10.1021/acsphotonics.5b00115} or in the long-time regime 
on the way towards the steady state.\cite{Gudmundsson16:AdP_10} 

Here, we present results concentrating on the charging current of, or the transport current
through, a system of two multilevel parallel quantum dots embedded in a short two-dimensional 
quantum wire placed in a photon cavity with one mode.
We use a recently developed Markovian version of the model to analyze the transport current
in the long-time limit in Liouville space.\cite{Gudmundsson16:AdP_10,2016arXiv161003223J}
Alternative scheme has been employed by Marino and Diehl that transform a Markovian
master equation into a dynamical field theoretical model that they subject to
a Keldysh functional renormalization to investigate nonequilibrium phase transitions in driven open 
systems.\cite{PhysRevB.94.085150} 

As we are describing transport through a multilevel system including both the 
para- and diamagnetic part of the electron-photon interaction,\cite{PhysRevE.86.046701}
we will not limited our scope to couple any particular two levels of the electron system resonantly with the 
cavity photon mode, but we will focus our attention on analyzing all underlying relaxation channels. 
Experiments on transport through electron systems in a photon cavity are usually not 
performed with time resolution, but in externally photon pumped cavities.\cite{PhysRevX.6.021014}
Calculation results from time resolved models of the underlying processes can serve as a basis for
understanding and interpreting experimental results with growing complexity of the systems,
as will be discussed below.

Our goal is to present a method to describe complex nanoscale electron systems placed in a photon cavity, 
which require a special attention to their geometry and all interactions and 
couplings present. We use a model with a state-dependent coupling tensor between the central electron
system and the semi-infinite external leads. Building the interacting equation of motion requires a stepwise introduction of 
the model complexities (interactions and components) together with appropriate truncation of the ensuing many-body spaces
at each step.\cite{Gudmundsson:2013.305,doi:10.1021/acsphotonics.5b00115}

\section{The model}
We consider a short two-dimensional quantum wire, incorporating two
parallel quantum dots, which is placed in a photon cavity. 
We will call this device the central system. 
The short quantum wire, and indirectly the entire central system, 
is weakly coupled to two external leads, the left (L), and the right (R)
leads, acting as electron reservoirs. The coupling opens up the central
system to electron transport through it at time $t=0$.
The model is appropriate for the description of weak coupling of the 
central system and the leads in the regime of sequential tunneling. 
We consider the leads to be at temperature $T=0.5$ K before the 
coupling, and assume this temperature to be higher than the 
Kondo temperature of the leads.
We do thus not describe processes in the Kondo regime.\cite{Kondo64:37,Lobo10:274007}

Our choice of a central system stems from the fact that with appropriate parameters
we have a system with a very rich character. The states of the system are
anisotropic and thus couple differently to different polarizations of the 
cavity photon field. The system has ``bound'' states with one or two electrons 
localized away from the contact area of the external leads. It has localized 
states in a quasi-continuum. Due to the anisotropy and the geometric information contained
in the model we have different selection rules for different relaxation channels, or
transitions. All these are properties that should appeal to researchers working in the 
experimental field. 

Below we first establish details of the central system of strongly coupled
electrons and photons and review its properties.
Then we describe the master equation formalism
used to model the time-dependent electron transport through the open
central system.

\subsection{The central system}
\begin{figure}[htb]
      \includegraphics[width=0.45\textwidth]{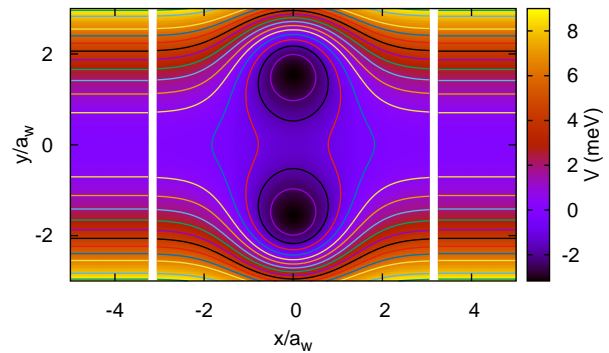}
      \caption{The central system connected to the parabolically confined semi-infinite external leads. 
               The common confinement energy of the leads and the central system $\hbar\Omega_0
               =2.0$ meV, and the small perpendicular external magnetic field $B=0.1$ T define
               a characteristic length $a_w=23.8$ nm. The white vertical gaps indicate the boundary between the 
               leads and the central system.}
      \label{Fig00}
\end{figure}

We consider a short two-dimensional quantum wire with two parallel quantum dots placed in a photon cavity.
The potential energy landscape defining the whole central system (the dots and the finite wire),
and the plunger gate voltage, $V_g$, used to raise or lower the central system with respect to the 
chemical potentials of the external leads, can be described by the formula
\begin{align}
      V(x,y) =& \left[\frac{1}{2}m^*\Omega_0^2y^2 +eV_g\right.\nonumber\\
             +& \left. V_d\sum_{i=1}^2\exp{\left\{-(\beta x)^2+\beta^2(y-d_i)^2\right\}} \right]\nonumber\\
             \times&\theta\left(\frac{L_x}{2}-x\right)\theta\left(\frac{L_x}{2}+x\right),
\label{Potential}
\end{align}
with $\hbar\Omega_0 = 2.0$ meV, $V_d = -6.5$ meV, $\beta = 0.03$ nm$^{-1}$, 
$d_1=-50$ nm, $d_2=+50$ nm, $L_x = 150$ nm, and $\theta$ is the Heaviside 
step function.  The potential is shown in Fig.\ \ref{Fig00}. The first line in Eq.\ (\ref{Potential})
describes the parabolic confinement of the short quantum wire in the $y$-direction, perpendicular
to the transport direction. 
The second line defines the potential of the quantum dots, and the third line
specifies the spatial domain of the short wire, beyond which we consider the potential infinite, supplying
the wire with hard wall ends in the $x$-direction.

Considering the small size of the system and the low bias to be used, 
the central system can contain several photons (0-16) and few electrons (0-3)
and is best described by a many-body formalism. The interactions of its constituents
are accounted for via a configuration interaction (CI) approach 
(also known as exact numerical diagonalization).\cite{Gudmundsson:2013.305} 
We assume GaAs parameters, $\kappa =12.4$, $m^*=0.067m_e$, and $g^*=-0.44$.
The parameters of the potential (\ref{Potential}) are chosen to investigate a
character rich GaAs system offering properties listed in the second paragraph of
this section, a system that is still feasible to attack numerically. 

The Hamilton operator for the central system in terms of the field operators is
\begin{align}
      H_\mathrm{S} =& \int d^2r \psi^\dagger (\mathbf{r})\left\{\frac{\pi^2}{2m^*}+
        V(\mathbf{r})\right\}\psi (\mathbf{r})
        + H_\mathrm{EM} + H_\mathrm{Coul}\nonumber\\ 
        -&\frac{1}{c}\int d^2r\;\mathbf{j}(\mathbf{r})\cdot\mathbf{A}_\gamma
        -\frac{e}{2m^*c^2}\int d^2r\;\rho(\mathbf{r}) A_\gamma^2,
\label{Hclosed}
\end{align}
with 
\begin{equation}
      {\bm{\pi}}=\left(\mathbf{p}+\frac{e}{c}\mathbf{A}_{\mathrm{ext}}\right),
\end{equation}
where $\mathbf{A}_{\mathrm{ext}}$ is a classical vector potential generating an external homogeneous
small magnetic field $B=0.1$ T along the $z$-axis perpendicular to the plane of the two-dimensional 
semiconductor, inserted in order to break spin and orbital degeneracies to enhance accuracy of the results. 
The external magnetic field, $B=0.1$ T, and the parabolic confinement energy of the leads and the 
central system $\hbar\Omega_0=2.0$ meV, lead together with the cyclotron frequency 
$\omega_c=((eB)/(m^*c))$ to an effective characteristic frequency 
$\Omega_w=({\omega_c^2+\Omega_0^2})^{1/2}$ and an effective magnetic length $a_w=(\hbar /(m^*\Omega_w))^{1/2}$.
This characteristic length scale, $a_w$, is used to scale all variables with dimension length
in the calculations, and assumes approximately the value 23.8 nm for parameters selected here.
In terms of the creation and annihilation 
operators, $a^\dagger$ and $a$, the Hamiltonian for the single cavity photon mode is 
$H_\mathrm{EM}=\hbar\omega a^\dagger a$, with energy $\hbar\omega = 0.8$ meV,
corresponding to the wavelength of 1.55 mm in air.
The electron-electron static Coulomb repulsion is represented by $H_\mathrm{Coul}$ written in terms
of four field operators 
\begin{equation}
      H_\mathrm{Coul} = \frac{1}{2}\int d^2rd^2r' \psi^\dagger (\mathbf{r}) \psi^\dagger (\mathbf{r}')
                        V_{\mathrm{Coul}}(\mathbf{r}-\mathbf{r}')\psi (\mathbf{r}') \psi (\mathbf{r}) ,
\end{equation}
and a spatially dependent Coulomb kernel.\cite{Gudmundsson:2013.305} 
\begin{equation}
      V_{\mathrm{Coul}}(\mathbf{r}-\mathbf{r}') = \frac{e^2}{\kappa\sqrt{|\mathbf{r}-\mathbf{r}'|^2+\eta^2}},
\end{equation}
with a small regularizing parameter $\eta/a_w=3\times 10^{-7}$.   
The quantized vector potential
of the cavity photon field is $\mathbf{A}_\gamma$. The last two terms of the Hamiltonian (\ref{Hclosed})
stand for the para- and the diamagnetic electron-photon interactions, respectively, necessary since we
will consider photon energy that might, or might not be, close to a transition resonance between particular 
electron states.\cite{Jonasson2011:01,PhysRevE.86.046701} The charge and the charge-current density operators are 
\begin{equation}
      \rho       = -e\psi^\dagger\psi, \quad
      \mathbf{j} = -\frac{e}{2m^*}\left\{\psi^\dagger\left({\bm{\pi}}\psi\right)
                 +\left({\bm{\pi}}^*\psi^\dagger\right)\psi\right\}.
\end{equation}

We consider a rectangular photon cavity $(x,y,z)\in\{[-a_\mathrm{c}/2,a_\mathrm{c}/2]
\times [-a_\mathrm{c}/2,a_\mathrm{c}/2]\times [-d_\mathrm{c}/2,d_\mathrm{c}/2]\}$ 
with the short two-dimensional quantum wire centered in the $z=0$ plane. Using the Coulomb gauge the
polarization of the electric field is parallel to the transport
in the $x$-direction (with the unit vector $\mathbf{e}_x$) by selecting the TE$_{011}$ mode, 
or perpendicular (defined by the unit vector $\mathbf{e}_y$) by selecting the 
TE$_{101}$ mode. 
For the cavity photons, the two versions of the quantized vector potential are in a stacked notation 
expressed as
\begin{equation}
      \mathbf{A}_\gamma (\mathbf{r})=\left({\hat{\mathbf{e}}_x \atop \hat{\mathbf{e}}_y}\right)
      {\cal A}\left\{a+a^{\dagger}\right\}
      \left({\cos{\left(\frac{\pi y}{a_\mathrm{c}}\right)}\atop\cos{\left(\frac{\pi x}{a_\mathrm{c}}\right)}} \right)
      \cos{\left(\frac{\pi z}{d_\mathrm{c}}\right)},
\label{Cav-A}
\end{equation}
for the TE$_{011}$ and TE$_{101}$ modes, respectively. The strength of the vector potential, ${\cal A}$,
determines the coupling constant $g_\mathrm{EM} = e{\cal A}\Omega_wa_w/c$, here set to 0.05 meV, 
leaving a dimensionless polarization tensor
\begin{equation}
      g_{ij}^k = \frac{a_w}{2\hbar}\left\{\langle i|\hat{\mathbf{e}}_k\cdot\bm{\pi}|j\rangle + \mathrm{h.c.}\right\},  
\end{equation}
where $|i\rangle$ and $|j\rangle$ are single-electron states of the short two-dimensional quantum wire,
$k=x,\,\mbox{or}\,\, y$. 
Latin indices are used for the single-electron states, and Greek for the 
many body states to be described below.

In order to find the energy spectrum and the states of the closed central system we use a stepwise scheme of
exact numerical diagonalizations and truncations:\cite{Gudmundsson:2013.305} First, the single-electron states
of the system are used to build a Fock space for 0-3 noninteracting electrons $\{|\mu\rangle\}$. This space is truncated
with respect to energy well above the bias window. For the parameters here we construct the Fock space with
the vacuum state, 36 one-electron states with energy up to 7.7 meV, 630 two-electron states, and 16 three-electron states.
In each sector of the Fock space the states lowest in energy are selected, and due to size of the central
system and the Coulomb repulsion energy we can neglect all many-body states with higher number of electrons. 
Second, the Hamiltonian of the Coulomb interacting electrons is diagonalized 
in the Fock basis creating a new Fock space of Coulomb interacting electrons $\{|\mu )\}$. Third, 
the 120 lowest in energy states in the Fock space of the Coulomb-interacting electrons are used to build a many-body basis  
for Coulomb interacting electrons and cavity photons by a tensor product of $\{|\mu )\}$ and the 17 lowest in energy eigenstates $|N\rangle$
of the photon number operator. This new basis is used to diagonalize the full Hamiltonian
of the central system (\ref{Hclosed}) obtaining a new Fock space of interacting electrons and photons $\{|\breve{\mu} )\}$,
with cavity-photon dressed electron states. 

As the numerical diagonalizations go beyond any simple order of perturbation calculations
this construction is reminiscent and parallel to
the possible construction of a Green function for the closed system (not done here), where one starts with a noninteracting Green function, and 
in a stepwise fashion first dresses it with the Coulomb interaction, and subsequently with the photon interactions through the Dyson equation
in both cases. The construction here is entirely carried out in functional spaces in a grid-free manner. 
Proceeding in this way either constructing an interacting Hamiltonian or a Green functions matrix is practically limited by
the size of the many-body space needed, be it for the diagonalization of the Hamiltonian or the solution of the Dyson equation. 
It is proper to stress again that in our case the small size of the system, the inherent energy scales and the strength of the 
Coulomb repulsion facilitate this task.

\begin{figure}[htb]
      \includegraphics[width=0.48\textwidth]{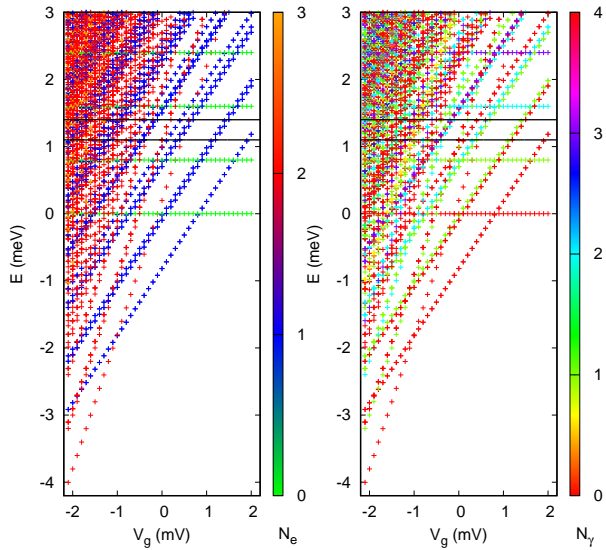}
      \caption{Energy spectrum for $x$-polarized cavity photons versus plunger gate voltage $V_g$. Electron number (left panel),
               and the mean photon number (right panel) is color coded into the states. 
               The horizontal black lines indicate the chemical potentials of the left, $\mu_L=1.4$ meV,
               and the right lead, $\mu_R=1.1$ meV, that will be coupled to the central system below.
               $\hbar\omega =0.8$ meV, and $g_\mathrm{EM}=0.05$ meV.}
      \label{Fig-Erof-x}
\end{figure}

\begin{figure}[htb]
      \includegraphics[width=0.48\textwidth]{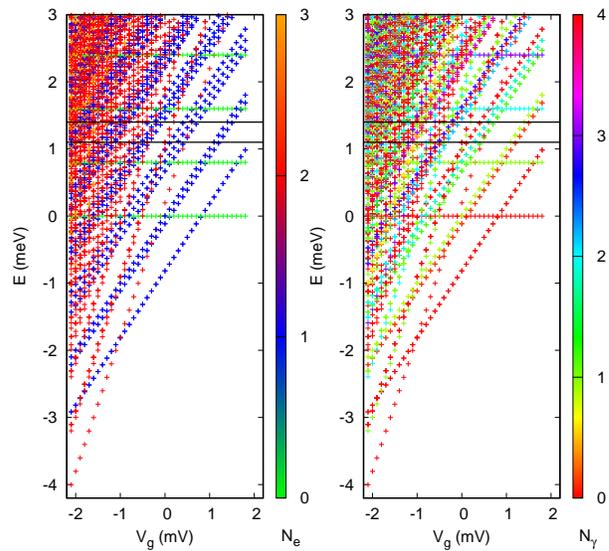}
      \caption{Energy spectrum for $y$-polarized cavity photons versus plunger gate voltage $V_g$. Electron number (left panel),
               and the mean photon number (right panel) is color coded into the states. 
               The horizontal black lines indicate the chemical potentials of the left, $\mu_L=1.4$ meV,
               and the right lead, $\mu_R=1.1$ meV, that will be coupled to the central system below.
               $\hbar\omega =0.8$ meV, and $g_\mathrm{EM}=0.05$ meV.}
      \label{Fig-Erof-y}
\end{figure}

The many-body energy spectra versus the plunger gate voltage $V_g$ for the central system are presented 
in Fig.\ \ref{Fig-Erof-x} for the $x$-polarization of the cavity photon field,
and in Fig.\ \ref{Fig-Erof-y} for the $y$-polarization.
In both cases we indicate with a color coding the electron content (an integer in the closed system) in the left panel of the figure,
and the photon number in the right panel. Without the electron-photon interaction the photon number is an integer, but the 
interaction does not conserve the number of photons. We will come back to this fact below. The dressed electron states 
contain in general are a linear combination of photon states, and  do thus
not represent an exact photon number and we can not refer to one- or two-photon replicas of an electron state, but as the replica
concept is useful we refer to the first or second replica of a certain electron state. 

In Figs.\ \ref{Fig-Erof-x} and \ref{Fig-Erof-y} we see a slight Rabi splitting of the first excitation of the 
one-electron ground state. The photon energy is 0.8 meV, but the separation of these two lowest one-electron 
states is around 0.72 meV. This ``detuning'' leads to the two branches of the Rabi-split states having a different
photon content. The Rabi splitting is larger for the $y$-polarization due to the anisotropy of the system and 
its states. 

For our parameters here we use 16 photon states. In our experience the convergence with respect to 
the photon basis is not critical, but instead the number of electron states is, as the photon dressing in the
strong coupling regime leads to an increased polarization or spreading of the charge of a 
state.\cite{ANDP:ANDP201500298}

\subsection{Coupling to the leads -- time evolution}
The central system is weakly coupled to the external leads and opened up
to electron transport through it at time $t=0$.
The semi-infinite leads are parabolically confined and are in the same external perpendicular classical weak magnetic field
as the central system (see Ref.\ \onlinecite{ThorstenPhD}, Sec.\ 3.4 and App.\ C, for an analytical calculation of their energy 
spectrum). In our calculations we include 4 of their subbands. The weak coupling is described by the 
Hamiltonian\cite{Moldoveanu09:073019,Gudmundsson09:113007}
\begin{equation}
      H_\mathrm{T}(t)=\sum_{i,l}\chi (t)\int dq\;
      \left\{T^l_{qi}c_{ql}^\dagger d_i + (T^l_{qi})^*d_i^\dagger c_{ql} \right\},
\label{H_T}
\end{equation}
where $d_i$ is the annihilation operator for an electron in the single-electron state of the 
the central system labeled with $i$, and $c_{ql}^\dagger$ is the creation operator
for an electron in lead $l$ with momentum and subband index labeled by the composite index 
$q$. The coupling tensor $T^l_{qi}$ describes the coupling
between these single-electron states of lead $l$ and the central system, and depends on the geometrical
form of the corresponding wave functions in the contact area extending approximately $a_w$ into each
subsystem.\cite{Gudmundsson09:113007,Gudmundsson:2013.305}
The remaining overall coupling constant to the leads is $g_{\mathrm{LR}}a_w^{3/2}=0.124$ meV. 
The switching of the coupling is determined by the Heaviside unit step function 
of time, at $t=0$.
Before the coupling, the electron system in the leads is at temperature $T=0.5$ K, and there exist no correlations
between the leads, or the leads and the central system.

The time evolution after
the coupling of the leads and the central system can be described by the Liouville-von Neumann equation for the
density operator of the system
\begin{equation}
    \partial_t \rho = \mathcal{L}\rho ,
\end{equation}
were, \( \mathcal{L}\) is the Liouville operator, defined by the commutator
\begin{equation}
    \mathcal{L}\rho = -i/\hbar\left[H, \rho\right],
\end{equation}
with $\rho$ the density operator of the total system, describing the dynamic state of both the leads and the central
system. As the energy spectra of the leads (the electron reservoirs) are dense, we have in earlier publications 
resorted to applying a formalism of Nakajima\cite{Nakajima58:948} and Zwanzig\cite{Zwanzig60:1338}
in which the dynamics of the whole system is projected on the central system leading to a generalized
master equation (GME)\cite{ANDP:ANDP201500298} 
\begin{equation}
      \partial_t{\rho_\mathrm{S}}(t) = -\frac{i}{\hbar}[H_\mathrm{S},\rho_\mathrm{S}(t)]
      -\frac{1}{\hbar}\int_0^t dt' K[t,t-t';\rho_\mathrm{S}(t')] 
\label{GME}
\end{equation}
for the reduced density operator $\rho_\mathrm{S}(t)$ describing properties of the central system under
influence of the external leads, and defined by tracing out variables of the leads 
$\rho_\mathrm{S}(t)=\mathrm{Tr_{LR}}\{\rho (t)\}$. The dissipative integral kernel, $K$, is constructed using terms 
of the coupling Hamiltonian (\ref{H_T}) up to second order, but the integro-differential form of the GME guarantees
higher order terms in its solution of the elemental type already present in $K$. 
In addition, the time-nonlocal structure of Eq.\ (\ref{GME}) brings in higher order terms
not present in an equivalent Markovian equation.\cite{Goan16:032113}

The solution to Eq.\ (\ref{GME}) can be found by numerical integration and 
iterations.\cite{Moldoveanu09:073019,Gudmundsson09:113007} However, for the $N=120$ states needed here 
for the transport calculations in the photon dressed
basis $\{|\breve{\mu})\}$, it is not feasible to integrate the GME in time much farther than 1000 ps. 
To go beyond that we have made a Markovian approximation to the GME (\ref{GME}) avoiding any further
approximations and using the Kronecker tensor product and vectorization of matrices to map the equation
from the Fock space of states to Liouville space of transitions.\cite{Gudmundsson16:AdP_10,2016arXiv161003223J}

In the $N^2$-dimensional Liouville space\cite{Weidlich71:325} the linear equation of motion 
is\cite{Nakano2010,Petrosky01032010}
\begin{equation}\label{eq:finalpde}
    \partial_t \vctrz(\rho_\mathrm{S}) = \mathfrak{L} \vctrz(\rho_\mathrm{S}),
\end{equation}
with $\mathfrak{L}$ being a general non-Hermitian operator with complex eigenvalues.
The vectorization of a matrix is accomplished by stacking its columns into a vector.\cite{2016arXiv161003223J}
The exact solution of Eq.\ (\ref{eq:finalpde}) is\cite{PhysRevB.81.155303}
\begin{equation}\label{eq:solution}
    \vctrz(\rho_\mathrm{S}(t)) =
    \left\{\mathfrak{U}\left[\exp{\left(\mathfrak{L}_\text{diag}t\right)}\right]\mathfrak{B}\right\}\vctrz(\rho_\mathrm{S}(0)),
\end{equation}
in terms of the left, $\mathfrak{U}$, and right, $\mathfrak{V}$, eigenvectors of $\mathfrak{L}$, 
\begin{align}
    \mathfrak{LB} &= \mathfrak{B L}_\text{diag},\nonumber\\
    \mathfrak{UL} &= \mathfrak{L}_\text{diag}\mathfrak{U},\\
    \intertext{with the normalization condition}
    \mathfrak{UB} &= I\nonumber\\
    \mathfrak{BU} &= I.
\end{align}
As the terms in $\mathfrak{L}$ can be cleanly traced back to the left or the right lead with no mixed terms we can,
as before,\cite{Moldoveanu09:073019,Gudmundsson09:113007} calculate the current from the left lead into the central
system, $I_L$, and the current from the central system into the right lead, $I_R$, using the time-derivative of 
$\vctrz(\rho_\mathrm{S}(t))$ in the equation of motion (\ref{eq:finalpde}) and its solution (\ref{eq:solution}).

Accuracy in numerical calculations is essential and error can cumulate in different parts of a model evaluation.
We have taken care designing all functional bases and necessary truncations thereof balancing 
RAM memory requirements and exactness of the results for the closed system within reasonable bounds, not 
shying away from heavy RAM usage (64 GB). This is only feasible with heavy parallelization of the code 
(using OpenMP on 24- and 32-core shared memory machines). 
As we use no time-integration of the equation of motion for the long 
times we are interested in, we escape errors seeping in through finite time steps in an integration. 
Instead, we have to pay attention to the quality of the complex valued eigenvalues of the nonsymmetric Liouvillian
${\cal L}$ in Eq.\ (\ref{eq:solution}). This is a nontrivial task and has been studied within the 
QuTiP python-framework.\cite{PhysRevE.91.013307,Johansson20131234} In our approach the accuracy of the 
eigenvalues and eigenvectors measured by the trace of the reduced density operator is such that the trace
is always equal to 1 with at least 8 accurate figures for all the cases described here, evaluated with
double precision in FORTRAN. 

Only 84 points, exponentially distributed on the time-axis, are used in the figures to conserve computational time
(900 CPU-hr on Intel Xeon CPU E5-2698 with 75\% parallelization efficiency).
A careful study of the details of the relaxation of the spin components forming a braided pattern, 
for example, to be presented in Fig.\ \ref{Fig_occ-Vg20} in the range $300<t<10000$ ps, would require more points.

\section{Results}
\begin{figure}[htb]
      \includegraphics[width=0.48\textwidth]{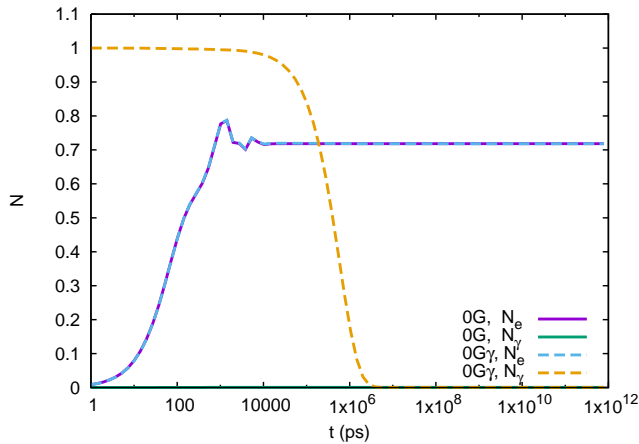}
      \caption{Mean electron and photon numbers in the central system as a function of time
               for initially the vacuum state (0G), and one photon in the system (0G$\gamma$).
               $x$-polarized photons, $V_g=+2.0$ mV, $B=0.1$ T, $\hbar\omega =0.8$ meV, and $g_\mathrm{EM}=0.05$ meV.}
      \label{Fig_Neg-Vg20}
\end{figure}

We explore the time-dependent electron transport into and through the system for three different
values of the plunger gate voltage, $V_g$. We start with $V_g=+2.0$ mV, when only the two spin components 
of the one-electron ground state are in the bias window, set by the chemical potentials of the left and 
right leads, $\mu_L=1.4$ meV and $\mu_R=1.1$ meV, respectively. We
expect resonant transport through the system. In Fig.\ \ref{Fig_Neg-Vg20} we show results for the mean electron
and photon number in the central system for two initial states, totally empty system (0G), and only one photon in 
the central system (0G$\gamma$).
In both cases the mean electron number rises quickly and remains constant just over 0.7. In the latter case (0G$\gamma$)
the mean photon number vanishes on a longer time scale, and for the former case the mean photon number never gains a value
visible in Fig.\ \ref{Fig_Neg-Vg20}.

\begin{figure}[htb]
      \includegraphics[width=0.48\textwidth]{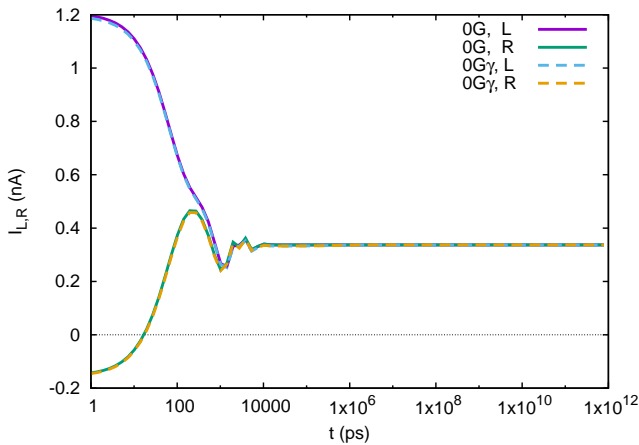}
      \caption{The current from the left lead (L), and into the right lead (R) as a function
               of time for initially the vacuum state (0G), and one photon in the system (0G$\gamma$).
               $x$-polarized photons, $V_g=+2.0$ mV, $B=0.1$ T, $\hbar\omega =0.8$ meV, and $g_\mathrm{EM}=0.05$ meV.}
      \label{Fig_curr-Vg20}
\end{figure}

Not surprisingly the current for the two cases, shown, in Fig.\ \ref{Fig_curr-Vg20}, is approximately 
the same.
For a short time the current is towards the central system from both leads (the right current, $I_R$, is negative in
the beginning). After 10 ns the system seems to be in a steady state with the current flowing through the 
central system. 

\begin{figure}[htb]
      \includegraphics[width=0.48\textwidth]{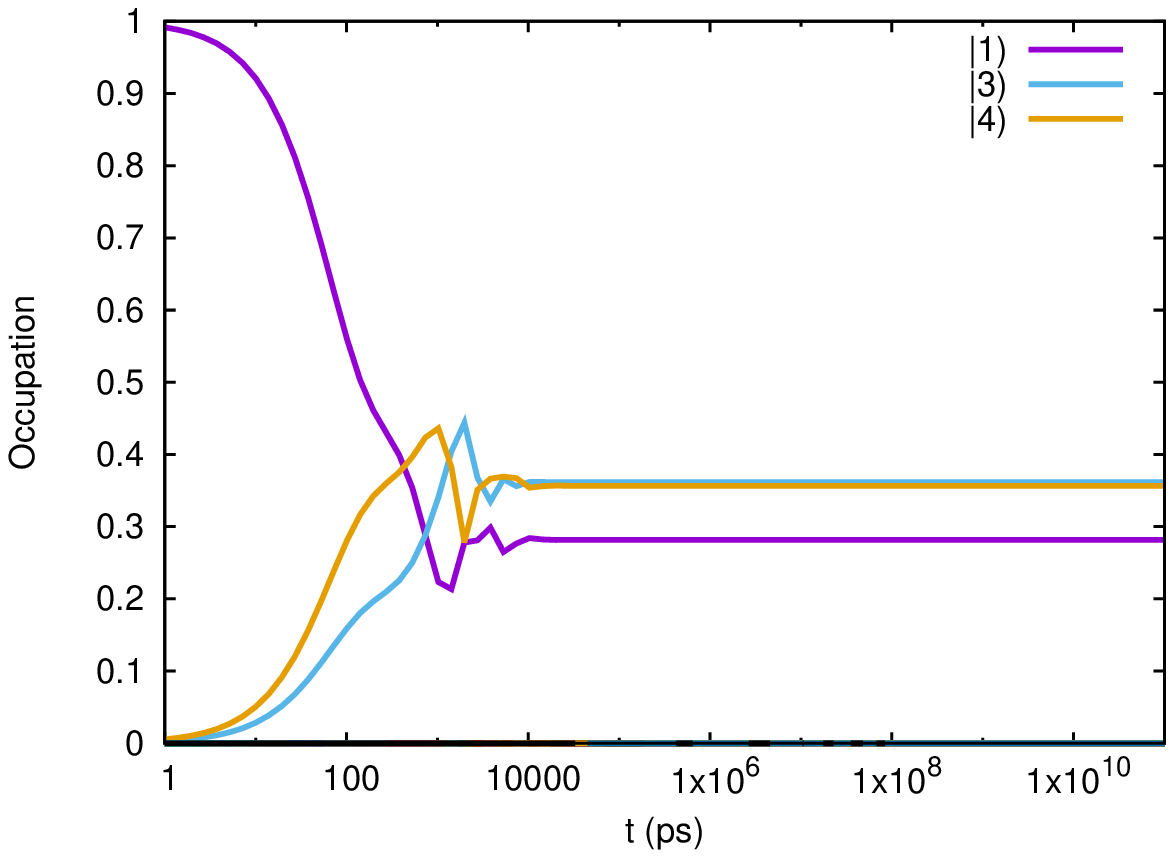}
      \includegraphics[width=0.48\textwidth]{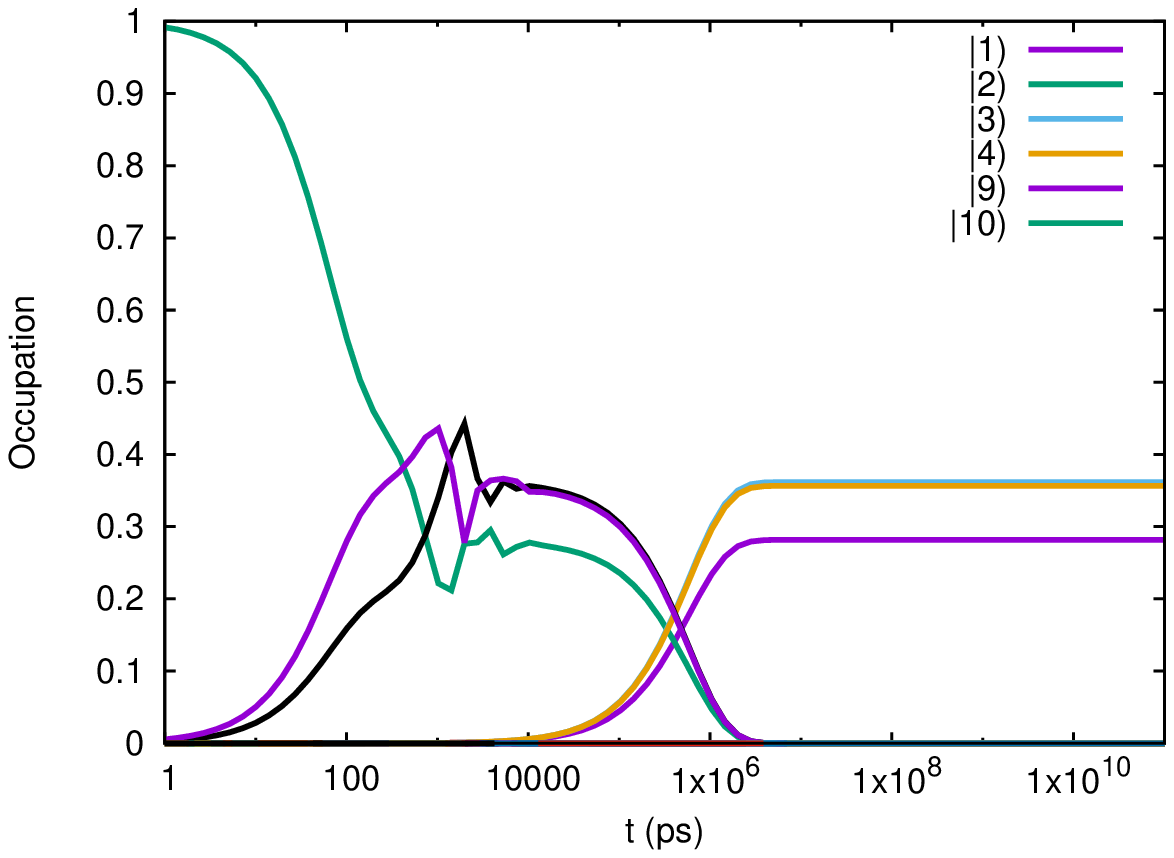}
      \caption{The time-dependent occupation of the states $|\breve{\mu})$ for initially the vacuum state (0G) (upper panel), 
               and one photon in the system (0G$\gamma$) (lower panel). $x$-polarized photons, 
               $V_g=+2.0$ mV, $B=0.1$ T, $\hbar\omega =0.8$ meV, and $g_\mathrm{EM}=0.05$ meV.}
      \label{Fig_occ-Vg20}
\end{figure}

This preliminary conclusion is not in agreement with the occupation of the many-body states of the central
system shown in Fig.\ \ref{Fig_occ-Vg20}.
In the upper panel of Fig.\ \ref{Fig_occ-Vg20}, we see for the former case (0G) that indeed the system starting 
in the vacuum state $|\breve{1})$ reaches
quickly a steady state formed by a mixture of $|\breve{1})$ and the partial population of the two spin components of the 
one-electron ground state, $|\breve{3})$ and $|\breve{4})$. The lower panel of Fig.\ \ref{Fig_occ-Vg20} shows that 
even when we see no changes in the current there are changes in the occupation of states. We see that due to the 
presence of one photon initially in the system charge is promoted temporarily to the intermediate states
$|\breve{8})$ and $|\breve{9})$, that are one-electron states with approximately one photon. They are photon replicas
of the two spin components of the one-electron ground state, and as such have energy 1.98 meV corresponding to 
one photon energy above the one-electron ground state in the bias window. We see thus, as noted earlier, 
photon associated tunneling into the system.\cite{Abdullah2014254} But, photons are not supplied to the 
system and the initial photon vanishes and the population transfers smoothly to the states 
$|\breve{3})$, $|\breve{4})$, and $|\breve{1})$ as before with no change seen in the current through
the central system. In the far-infrared (FIR) regime the electromagnetic transitions in the system are 
slow so the steady state is only reached after 10 $\mu$s. In addition, it is important to have in mind that 
electromagnetic transitions are only active due to the coupling to the leads, which is weak here.
Without this coupling the many-body states of the central system are its eigenstates.

\begin{figure}[htb]
      \includegraphics[width=0.48\textwidth]{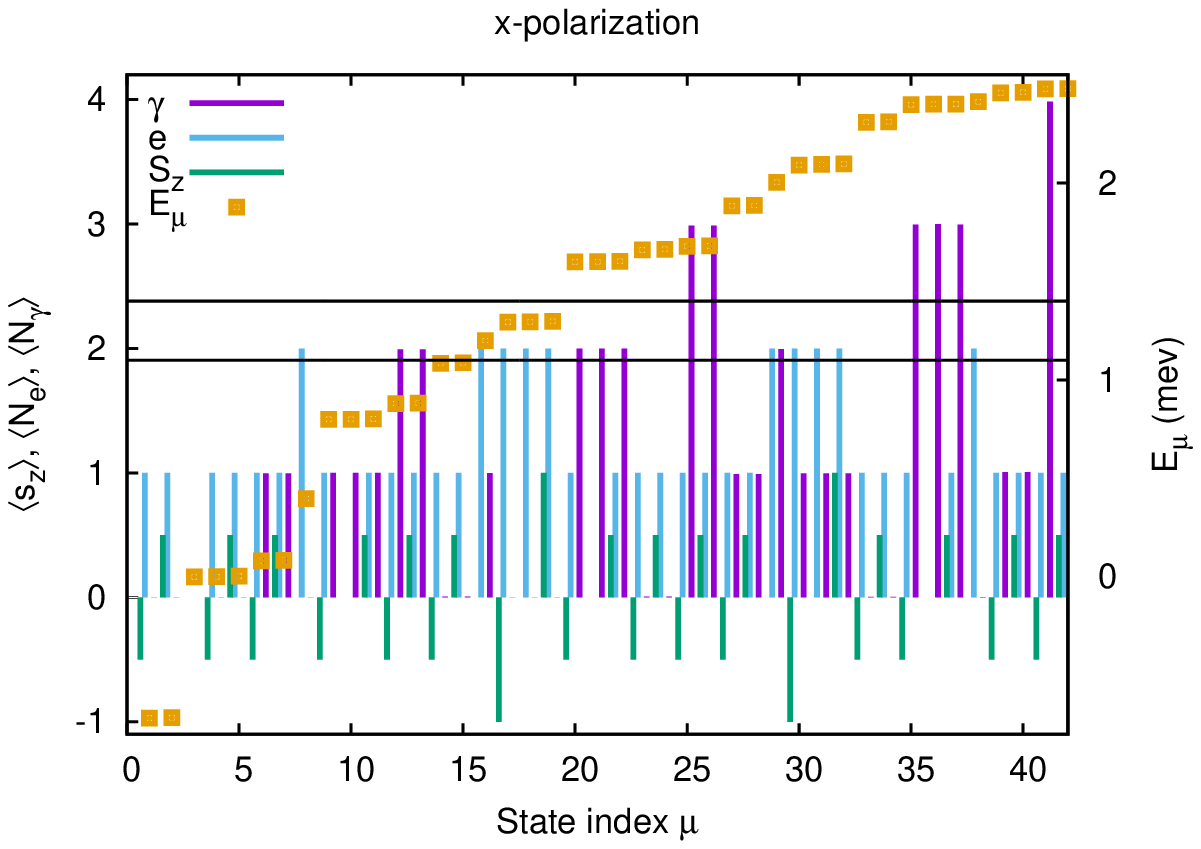}
      \includegraphics[width=0.48\textwidth]{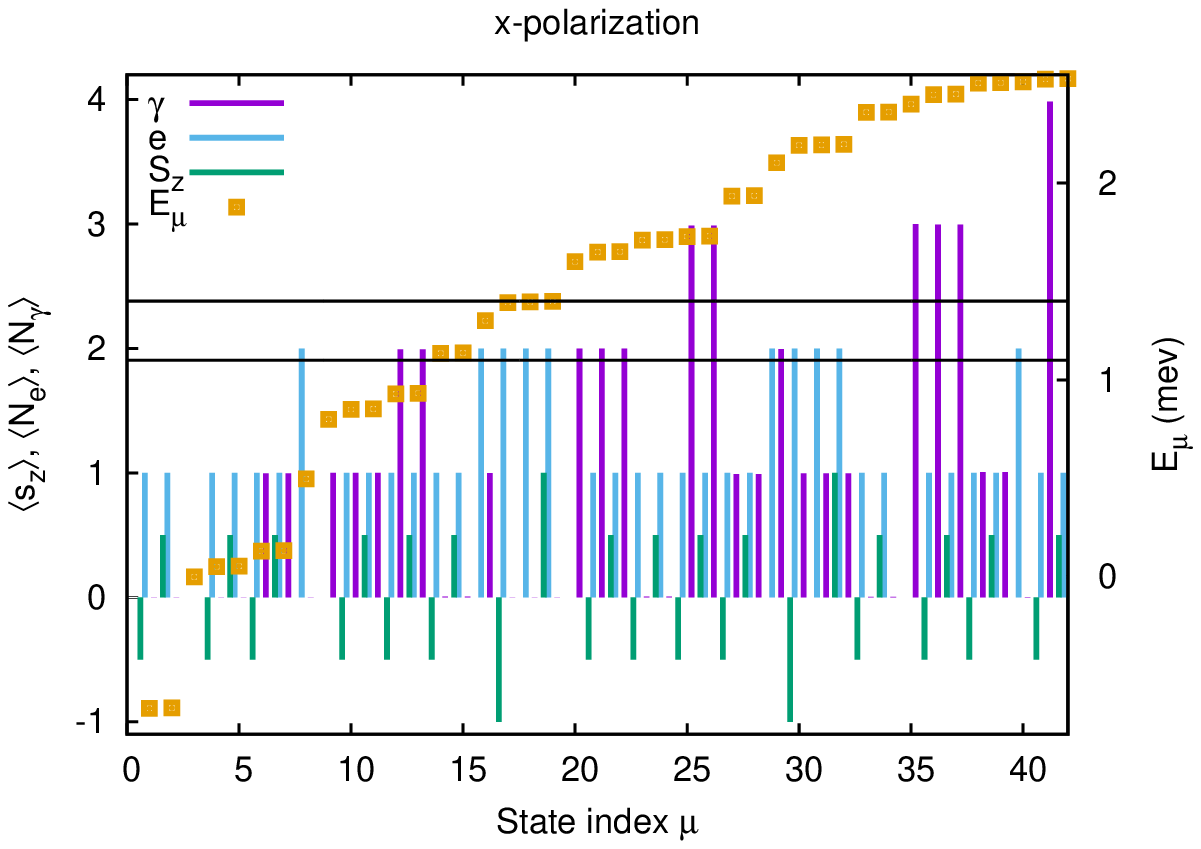}
      \caption{The energy spectrum (right $y$-axis), the mean electron, the 
               mean photon content, and the spin of the 42 lowest many-body states
               for $V_g=+0.1$ mV (upper panel), and $V_g=+0.15$ mV (lower panel). 
               The upper horizontal black line marks the chemical potential of the 
               left lead $\mu_l$, and the lower one $\mu_r$, the chemical potential of
               the right lead. $x$-polarized photons.
               $\hbar\omega =0.8$ meV, and $g_\mathrm{EM}=0.05$ meV.}
      \label{Fig_ErofNeg}
\end{figure}

We now turn to more complex situations with the plunger gate voltage set at
$V_g=+0.1$ and $+0.15$ mV shown in Fig.\ \ref{Fig_ErofNeg}.
For the case of $V_g=+0.1$ mV, the upper panel of Fig.\ \ref{Fig_ErofNeg}, we see a
photon replica of the singlet two-electron ground state together with the three spin components of the 
lowest two-electron triplet state. Just below the bias window are two one-electron states with 
vanishing photon components. In the lower panel of Fig.\ \ref{Fig_ErofNeg} we have just
changed the plunger gate voltage to $+0.15$ mV in order to raise the one-electron
states into the bias window still keeping the four two-electron states there. 
We use this slight shift in the plunger gate voltage $V_g$ to
shed light on the roles assumed by the one- and two-electrons states in our
transport picture, built on sequential tunneling.

\begin{figure}[htb]
      \includegraphics[width=0.48\textwidth]{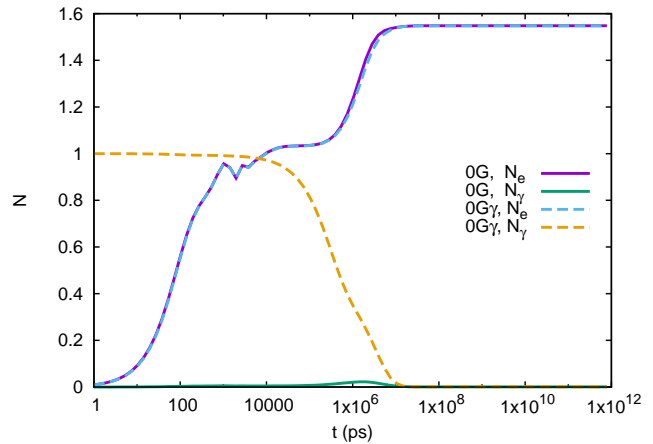}
      \caption{Mean electron and photon numbers in the central system as a function of time
               for initially the vacuum state (0G), and one photon in the system (0G$\gamma$).
               $x$-polarized photons, $V_g=+0.1$ mV, $B=0.1$ T, $\hbar\omega =0.8$ meV, and $g_\mathrm{EM}=0.05$ meV.}
      \label{Fig_Neg-Vg01-0Gx}
\end{figure}

Here, we have more obvious choices for the initial state of the system. In Fig.\ \ref{Fig_Neg-Vg01-0Gx}
we see for the case of an initially empty system (0G), or with only one photon initially (0G$\gamma$), 
that the photons seem not to play any major role. 
The system is charged quickly to one electron, and further delayed charging seems to go in hand with the 
vanishing of the initial photon in the system, or if there was none in the system initially, with the 
appearance of a very small photon component, that again vanishes. As there are several states available
below the bias window we might expect charge to get trapped in the system suppressing possible
current through it in the steady state.

\begin{figure}[htb]
      \includegraphics[width=0.48\textwidth]{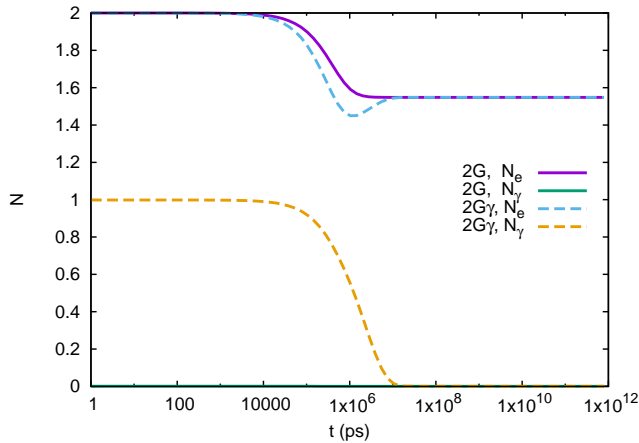}
      \caption{Mean electron and photon numbers in the central system as a function of time
               for initially the two-electron ground state (2G), and the first photon 
               replica of the two-electron ground state of the system (2G$\gamma$).
               $x$-polarized photons, $V_g=+0.1$ mV, $B=0.1$ T, $\hbar\omega =0.8$ meV, and $g_\mathrm{EM}=0.05$ meV.}
      \label{Fig_Neg-Vg01-2Gx}
\end{figure}

Before looking at the corresponding current and occupation, we present in Fig.\ \ref{Fig_Neg-Vg01-2Gx}
the evolution of the mean electron and photon numbers for the two electrons initially in the 
central system, in the two-electron ground state (2G), and the first photon replica (2G$\gamma$) thereof. 
Here, all changes are slow, occurring for time between 100 ns and 10 $\mu$s.
For the first time, we see here that the discharging of the system
depends on the initial photon number. It is faster for the first photon replica of the 
two-electron ground state (2G$\gamma$) than for the corresponding ground state (2G). We should have in 
mind that the $x$-polarized photon field in the system stretches the charge density into
the contact area of the central system facilitating the charge to leave or enter the 
system.\cite{0953-8984-28-37-375301}

\begin{figure}[htb]
      \includegraphics[width=0.48\textwidth]{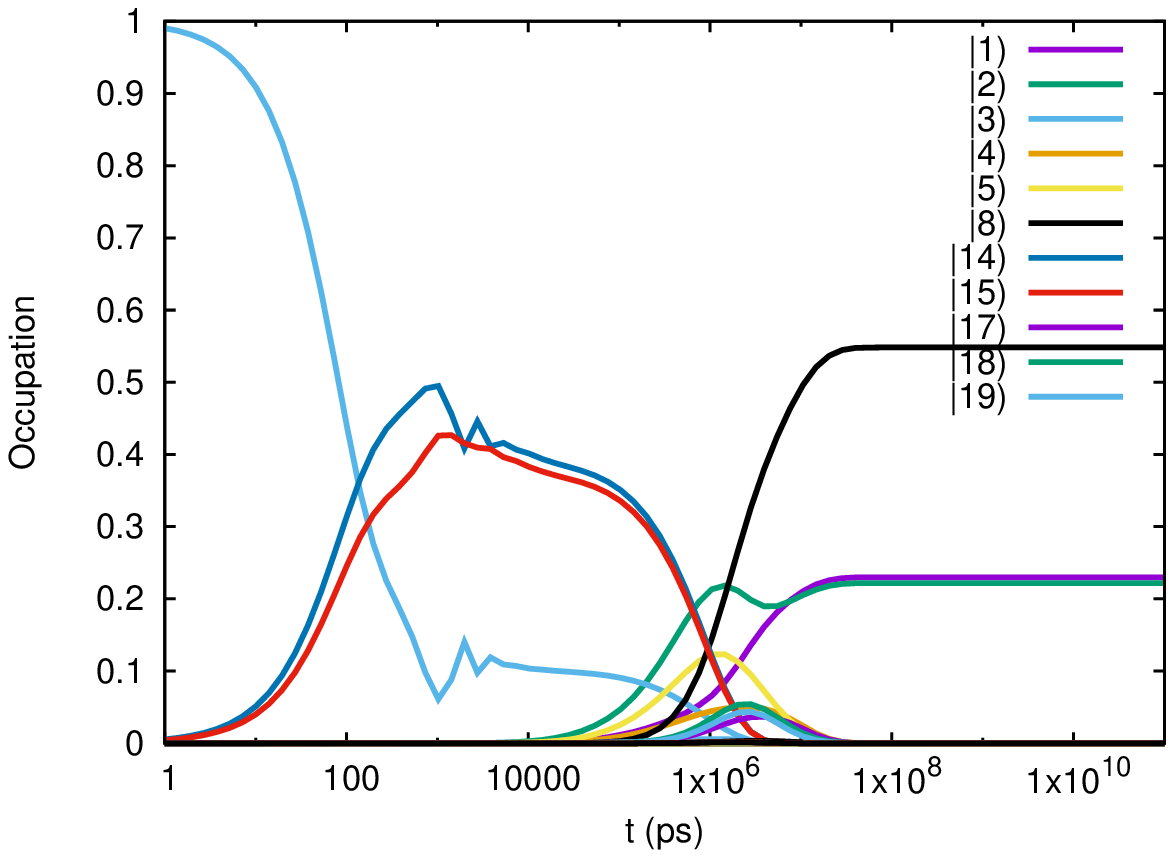}
      \includegraphics[width=0.48\textwidth]{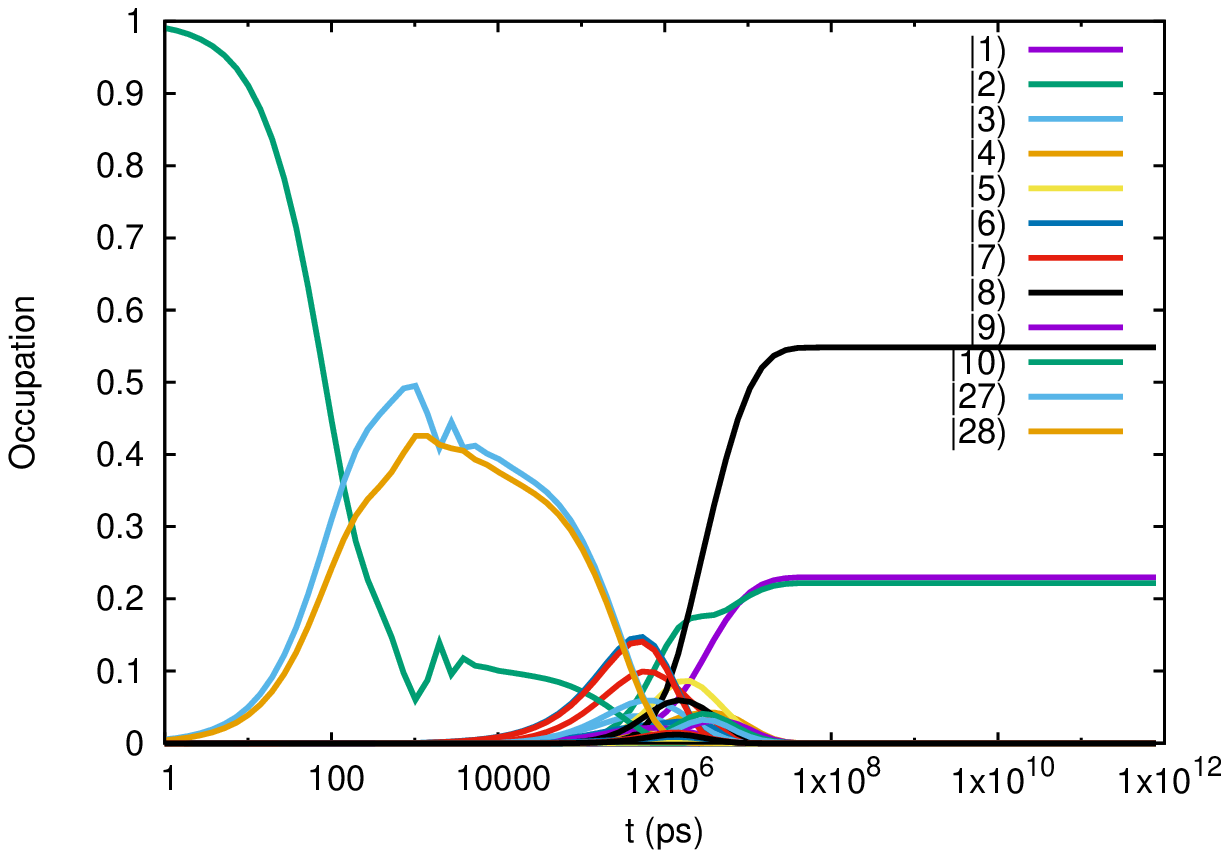}
      \caption{The time-dependent occupation of the states $|\breve{\mu})$ for initially the vacuum state (0G) (upper panel), 
               and one photon in the system (0G$\gamma$) (lower panel). $x$-polarized photons, 
               $V_g=+0.1$ mV, $B=0.1$ T, $\hbar\omega =0.8$ meV, and $g_\mathrm{EM}=0.05$ meV.}
      \label{occ-Vg01-0Gx}
\end{figure}

In Fig.\ \ref{occ-Vg01-0Gx} we present the time-dependent occupation of states
for the case of no electron initially in the system, but either no (0G) (upper panel), 
or one photon (0G$\gamma$) (lower panel), corresponding to Fig.\ \ref{Fig_Neg-Vg01-0Gx}.
For the (0G) case in the upper panel we identify the one-electron states $|\breve{14})$ and $|\breve{15})$
with energy $E\approx 1.08$ meV just below the bias window that are active in charging the system. The 
steady state is a combination of the two spin components of the one-electron ground state, and with
a higher probability the two-electron ground state. The intermediate states active around 3 $\mu$s are the 
two-electron triplet states $|\breve{17})$, $|\breve{18})$, and $|\breve{19})$ together with the one-electron
state $|\breve{5})$, all states with odd spatial parity (measured along the $y$-axis), that is slightly broken 
by the external magnetic field. As we use a state-dependent coupling to the leads that depends on the 
spatial properties of the wave functions, the coupling of these states to the lowest subband in the leads
with even parity is small. 

For the (0G$\gamma$) case displayed in the lower panel of Fig.\ \ref{Fig_Neg-Vg01-0Gx} we see the initial
one-photon state $|\breve{10})$ rapidly depopulated with the one-electron states $|\breve{27})$ and $|\breve{28})$
containing one photon and having energy 1.88 meV taking over. These states are the first photon replicas of the one-electron
states we noticed just below the bias window. Again, we are observing a photon associated charging of the system, and 
before the system reaches the same steady state as in the (0G) case we see intermediate states of two types;
low energy one-electron states with no photon content like $|\breve{4})$ and $|\breve{5})$,
and one-electron states with approximately one photon $|\breve{6})$ and $|\breve{7})$, and last we
see small occupation of the triplet states in the bias window. So, differently from the (0G) case we see radiative 
transitions on the way to the steady state.

\begin{figure}[htb]
      \includegraphics[width=0.48\textwidth]{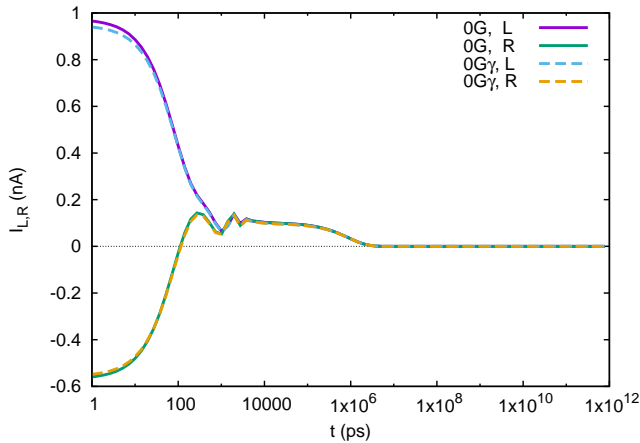}
      \caption{The current from the left lead (L), and into the right lead (R) as a function
               of time for initially the vacuum state (0G), and one photon in the system (0G$\gamma$).
               $x$-polarized photons, $V_g=+0.1$ mV, $B=0.1$ T, $\hbar\omega =0.8$ meV, and $g_\mathrm{EM}=0.05$ meV.}
      \label{current-Vg01-0Gx}
\end{figure}

Not surprisingly, we see in Fig.\ \ref{current-Vg01-0Gx} for $t<200$ ps high charging current into the 
system. This is followed by a considerable current through the system up to the time of 1 $\mu$s during which period 
it is relaxing on the its way to the steady state. But, the final and real steady state is not fully reached until 
after approximately 20 $\mu$s.  
The logarithmic time scale in the figures can be a bit deceiving when estimating the charge flowing
into and through the system. As can be confirmed by Fig.\ \ref{Fig_Neg-Vg01-0Gx} approximately one
electron enters the central system in the charging phase lasting to 200-500 ps, but in the intermediate
time regime extending to 1 $\mu$s the number of electrons passing through is of the order of 10$^5$, 
although on the average only 0.6 are staying in!

\begin{figure}[htb]
      \includegraphics[width=0.48\textwidth]{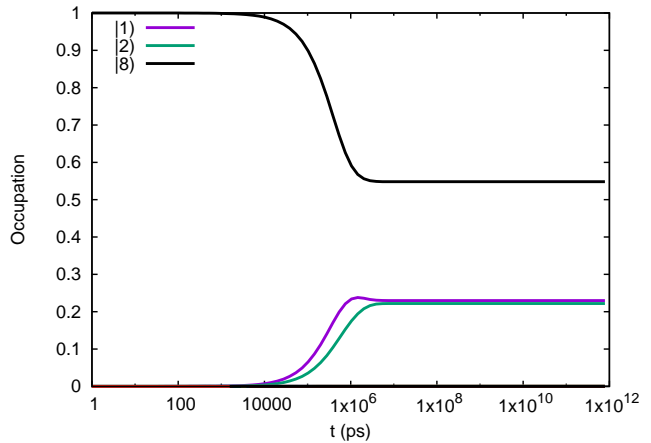}
      \includegraphics[width=0.48\textwidth]{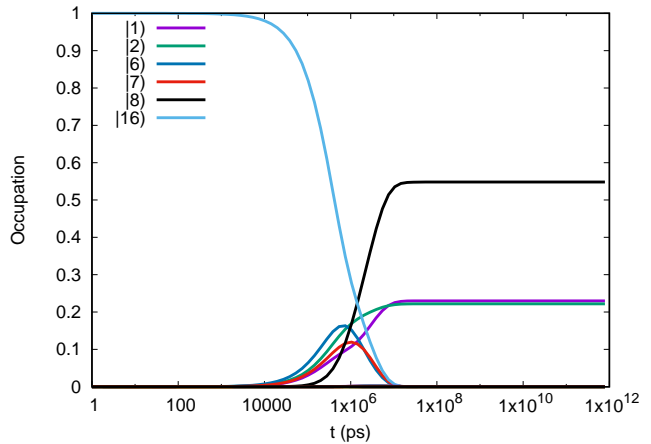}
      \caption{The time-dependent occupation of the states $|\breve{\mu})$ for initially the two-electron ground state (2G) (upper panel), 
               and the first photon replica of the two-electron ground state of the system (2G$\gamma$) (lower panel).
               $x$-polarized photons, $V_g=+0.1$ mV, $B=0.1$ T, $\hbar\omega =0.8$ meV, and $g_\mathrm{EM}=0.05$ meV.}
      \label{occ-Vg01-2Gx}
\end{figure}

Glancing at Figure\ \ref{current-Vg01-0Gx} one is tempted to conclude that there is
no steady state current through the system. In order to explore this situation we now analyze in more
detail what happens when initially there are two electrons in the system. We started with this in Fig.\
\ref{Fig_Neg-Vg01-2Gx}, but now turn to the occupation (population) for the (2G) and (2G$\gamma$) cases shown in
Fig.\ \ref{occ-Vg01-2Gx}.
The (2G) case with initially 2 electrons in the ground state is a simple case with no electromagnetic
transitions active. Very slowly the two-electron ground state $|\breve{8})$ (a singlet) looses charge to the 
two spin components of the one-electron ground state, $|\breve{1})$ and $|\breve{2})$. 
This transition is slow as the single-electron wave functions associated with these states have a small weight
in the contact area of the short wire in the central system, and all transitions are activated by the coupling
to the leads, that depends on the wave functions in that region. 

The relaxation of the charge from the first photon replica of the two-electron ground state $|\breve{16})$
(the 2G$\gamma$ case seen in the lower panel of Fig.\ \ref{occ-Vg01-2Gx}) in the bias window is faster with
of course the same steady state reached in the end, but now the active intermediate states are the one-electron
states $|\breve{6})$ and $|\breve{7})$, a radiative transition terminating in the two spin components of the 
one-electron ground state $|\breve{1})$ and $|\breve{2})$.

\begin{figure}[htb]
      \includegraphics[width=0.48\textwidth]{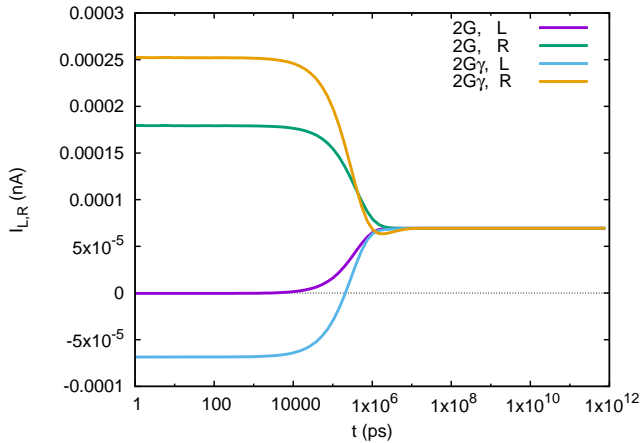}
      \caption{The current from the left lead (L), and into the right lead (R) as a function
               of time for initially the two-electron ground state (2G), 
               and the first photon replica of the two-electron ground state of the system (2G$\gamma$).
               $x$-polarized photons, $V_g=+0.1$ mV, $B=0.1$ T, $\hbar\omega =0.8$ meV, and $g_\mathrm{EM}=0.05$ meV. }
      \label{current-Vg01-2Gx}
\end{figure}

Now, we are prepared for viewing the current into and through the system for both the cases with
initially 2 electrons in the system (2G and 2G$\gamma$) displayed in Fig.\ \ref{current-Vg01-2Gx}.
For the long transient regime up to 1 $\mu$s we see the system is slowly loosing charge as 
$I_L$ is negative, indicating a current from the central system into the left lead, and $I_R$ 
is positive indicating current from the central system into the right lead.
But, we notice a \textit{nonzero} steady state current through the system, albeit small, 
that we have to analyze further. Before, we have to ask the question whether even in the 
case of an initially empty system (see Fig.\ \ref{current-Vg01-0Gx}) we also had this finite but small 
steady-state current without noticing it. 

\begin{figure}[htb]
      \includegraphics[width=0.48\textwidth]{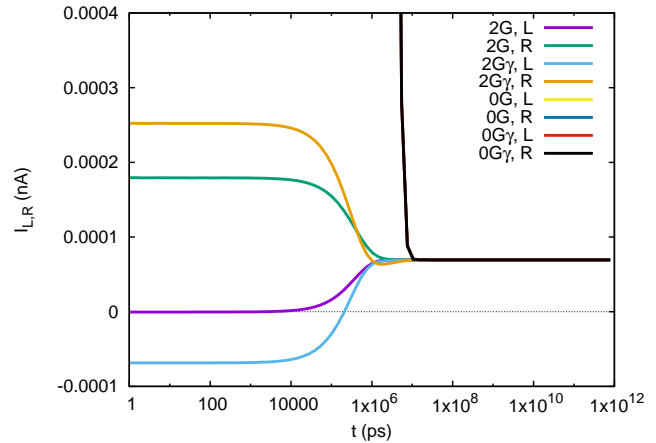}
      \includegraphics[width=0.48\textwidth]{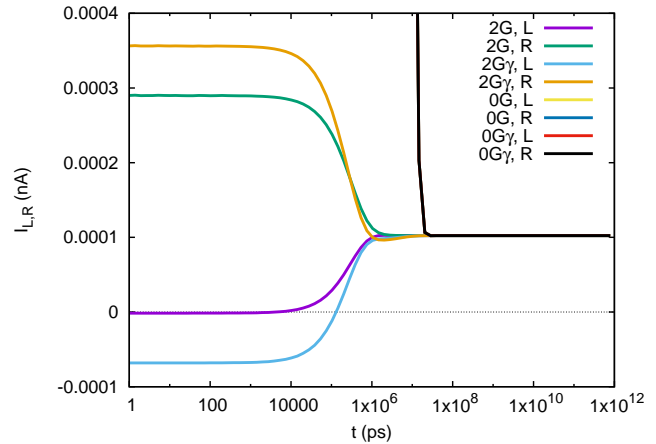}
      \caption{Comparison of the current from the left lead (L), and into the right lead (R) 
               as a function of time for several initial states. $V_g=+0.1$ mV (upper panel), 
               and $V_g=+0.15$ mV (lower panel), $x$-polarized photons, $B=0.1$ T, $\hbar\omega =0.8$ meV, 
               and $g_\mathrm{EM}=0.05$ meV.}
      \label{current-compare-x}
\end{figure}

Indeed, Fig.\ \ref{current-compare-x} confirms that there is the same 
steady-state current through the system for the different initial cases, independent of
whether we start with none, or two electrons in the system. 
Even, when we displace slightly the plunger gate voltage to $V_g=+0.15$ mV and thus lift the 
two one-electron states $|\breve{14})$ and $|\breve{15})$ into the bias window we continue to
have a slight steady state current through the system. Where does it come from? These two 
one-electron states are not populated in the steady state.

\begin{figure}[htb]
      \includegraphics[width=0.48\textwidth]{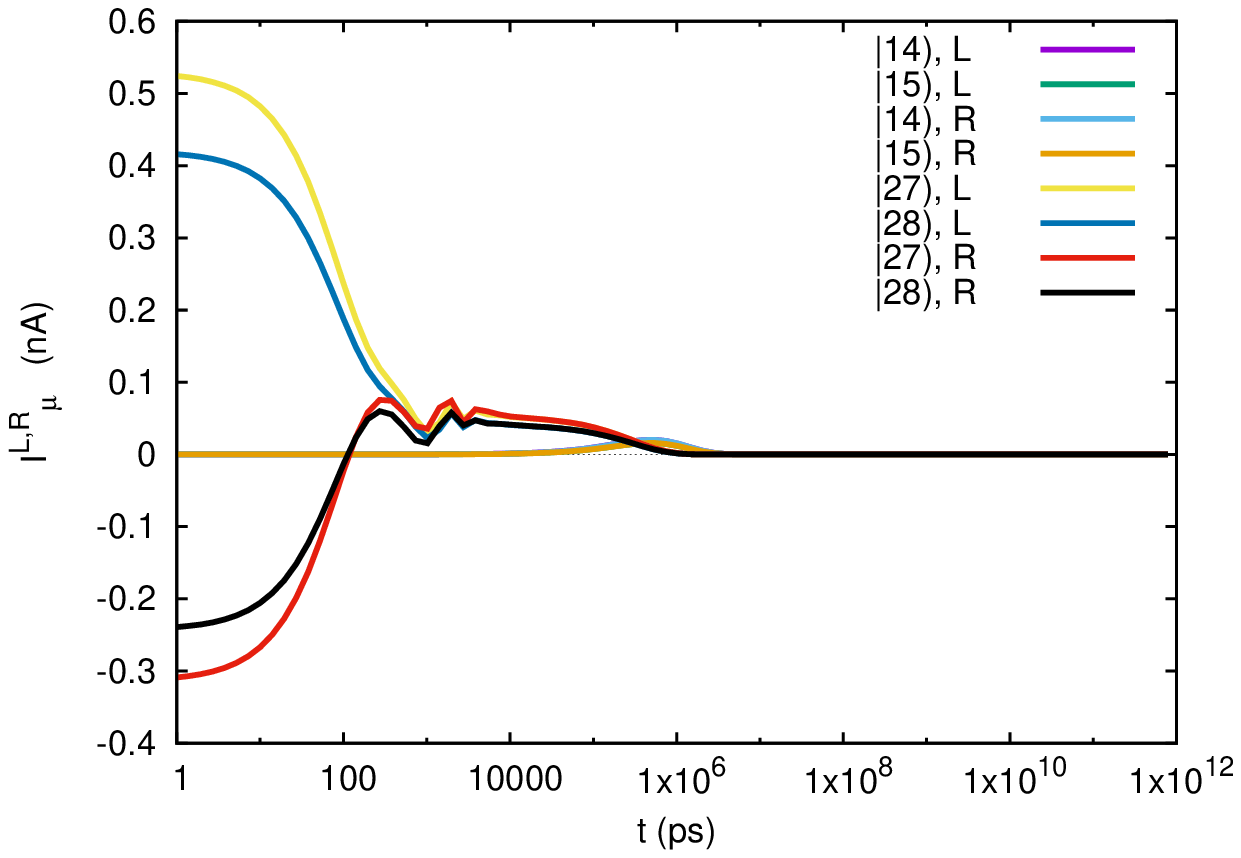}
      \includegraphics[width=0.48\textwidth]{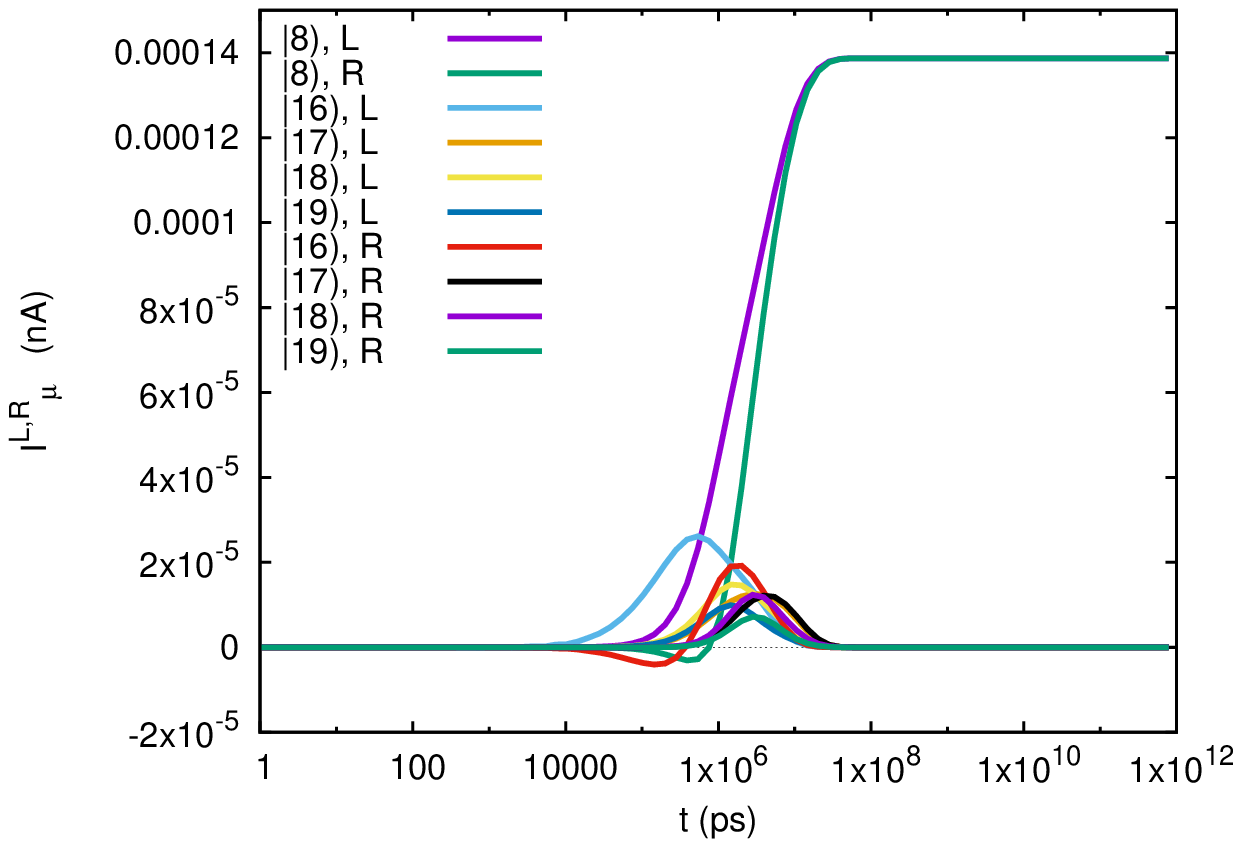}
      \caption{The partial current carried by the state $|\breve{\mu})$ from the left lead (L), and into the right lead (R)
               for the most active one-electron states (upper panel), and the most active
               two-electron states (lower panel). $x$-polarized photons, 
               $V_g=+0.1$ mV, $B=0.1$ T, $\hbar\omega =0.8$ meV, and $g_\mathrm{EM}=0.05$ meV.}
      \label{Irlmu-x}
\end{figure}

The answer can be found by analyzing the partial current through each state in the system,
a feature of the model that can not be repeated in experiments, but can give a valuable
insight into the underling active processes in the system.
In Fig.\ \ref{Irlmu-x} we display in the upper panel the contribution of the active one-electron
states in the system, and in the lower panel the active two-electron states. 
In addition, we have selected states in the bias window that could be expected to carry current.
In the upper panel we see that the one-electron states contribute to the charge loss from the 
system as we noticed earlier. The steady state current of these one-electron state seems to 
vanish. We have confirmed that this is indeed true.
In the lower panel of Fig.\ \ref{Irlmu-x} we notice that the steady state current is only carried by
the two-electron ground state $|\breve{8})$, even though it is not in the 
bias window. How is this possible?

As we saw earlier (see Fig.\ \ref{occ-Vg01-2Gx}) the two-electron ground state is the 
state with the highest energy contributing significantly to the steady state.
In the steady state the charge has all relaxed to states below the bias window.
The two-electron ground state (a spin singlet) with even parity (with respect to the $y$-direction) has 
higher coupling to the lowest subband of the leads than the spin triplet states in the bias window.
We are thus seeing an off-resonance inelastic (with respect to the system lead tunneling) contribution 
to the steady-state current by the two-electron ground state. The current is low, as it is inelastic 
and the coupling to the two-electron state mostly confined to the parallel quantum dots is low 
on grounds of the geometry.

Even though we can talk about a strong electron-photon coupling here, it should be stated that for
the $x$-polarized cavity field we are well off-resonance with the photon number close to an integer 
for most states. For the $y$-polarization there is a weak Rabi splitting for the low lying one-electron
states, but not for the two-electron states, and here we used two-electron states as the initial state
or started with no electrons in the central system.

\begin{figure}[htb]
      \includegraphics[width=0.48\textwidth]{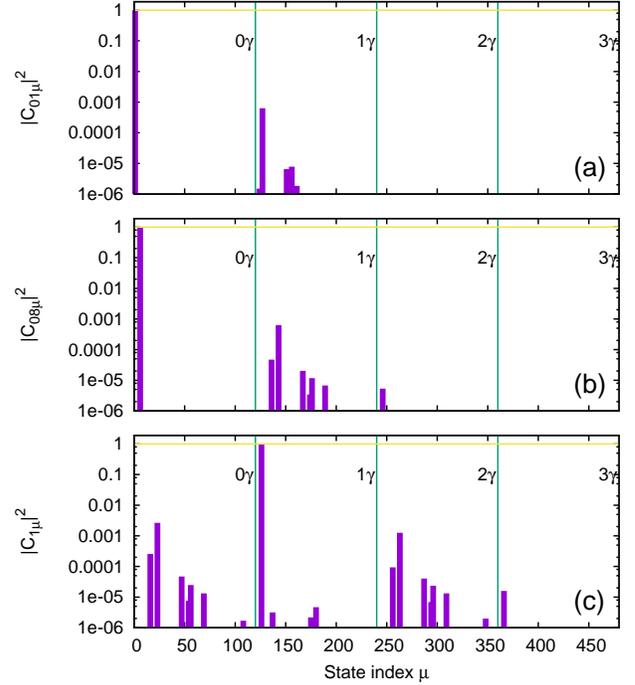}
      \caption{Spectral composition of the one-electron ground state $|\breve{1})$ (1G) (a),
               the two-electron ground state $|\breve{08})$ (2G) (b), and the first photon replica of the 
               two-electron ground state $|\breve{16})$ (2G$\gamma$) (c), in terms of a tensor product of 
               Coulomb interacting electron states $|\mu )$ and eigenstates $|n\rangle$ of the photon number operator.
               The photon sectors are marked by their photon number and indicated with vertical thin green lines. 
               The unit probability is indicated with a thin yellow horizontal line.
               $\hbar\omega =0.8$ meV, and $g_\mathrm{EM}=0.05$ meV.}
      \label{FigRes}
\end{figure}

The fact that electron-photon system is only in a weak resonance here can be confirmed by inspecting 
Figs.\ \ref{Fig-Erof-x}, \ref{Fig-Erof-y}, and \ref{Fig_ErofNeg}. All the same, it is necessary to 
have in mind that even for the $x$-polarization here, we have dressed electron states with a mixed photon content. 
This is shown in Fig.\ \ref{FigRes} for the one- and two-electron ground states, ($|\breve{1})$ and $|\breve{8})$), 
and the first replica of the two-electron ground state ($|\breve{16})$). 
In case of a stronger resonance between the electrons and the cavity photons, we have observed in an earlier
publication stronger influence of radiative transitions on the evolution of the system and stronger variations
in the mean photon number.\cite{Gudmundsson16:AdP_10,2016arXiv161003223J}

\section{Discussion}
We have used our model of electron transport through a central system with interacting electrons and 
photons to describe the time evolution of the system after it has been started/prepared in a certain eigenstate of the 
closed system. Time resolved experiments in this field have not yet been performed, but transport experiments 
on electron systems in photon pumped cavities are gaining strength.\cite{Delbecq11:01,PhysRevX.6.021014}
The importance of time-dependent modeling of the systems is twofold: First, it allows us to gain insight into
the underlying processes in the system. Second, with the knowledge of these active fundamental processes and their time
evolution we can research the states of the system we would like to maintain with an external photon pumping, as is
commonly done in experiments, even in a system with a complex structure.

As expected for a small bias, resonant tunneling through the lowest states of the central system quickly brings it into
a steady state. Judging from the transport current the relaxation seems similar for the case of initially one or no
photon in the cavity. The current is not sensitive to the initial number of photons, but a look at the 
occupation of states during the process shows, as could be anticipated, that of course different
states take part. The one photon initially in the system promotes the incoming electron to a
photon replica of the one-electron ground state located above the bias window. The following internal conversion, or
transition, is not seen in the current. Thus the time point when the system reaches the steady state does
depend on the initial photon number. The conclusion is that the transport current can not be used as an
indication of when the system reaches the steady state. (See, for example, Figs.\ \ref{Fig_curr-Vg20} and \ref{Fig_occ-Vg20}).

The situation becomes more complex if the bias window defined by the external leads 
is located higher in the many-body energy spectrum. Then the character of the states in the window and below it, 
shaped by their geometry and the coupling between the constituents of the system, are of importance. 
In these cases we find a small but important off-resonance current through the system. 

Even so, there is a common theme to the results for both cases. The upper panels of Figs.\ \ref{Fig_occ-Vg20}
and \ref{occ-Vg01-2Gx} display direct relaxation of the system to its steady state, while the lower panels of
Figs.\ \ref{Fig_occ-Vg20} and \ref{occ-Vg01-2Gx}, and both panels of Fig.\ \ref{occ-Vg01-0Gx} show higher
order processes with intermediate states. This can be compared to the Markovian time evolution of a simple
damped quantum harmonic oscillator linearly coupled to a reservoir with a vanishing expectation value for its excitation.
If it is initially placed in a high-energy state it has to cascade through all the lower (intermediate states)
on its path to the ground state, the stead state. The cascading is a signature of higher order effects in terms
of the system reservoir coupling. The twist to this picture is that in our model the reservoir is the external
leads, but not a photon reservoir, and in the Liouville space formulation the steady state, reflecting higher order processes,
can be found directly without a time integration revealing the history of the time-dependent occupation of the
intermediate states. Of course this information is hidden in the complex spectrum of the 
Liouville operator.\cite{2016arXiv161003223J} 

Just a reminder, as mentioned after the introduction of the non-Markovian (Eq.\ (\ref{GME}))
and the Markovian (Eq.\ (\ref{eq:finalpde})) equations of motion, there are higher order effects
associated with the non-Markovian equation (\ref{eq:finalpde}) that are not present in the 
Markovian one (\ref{GME}), caused by the nonlocal-time structure of the former.\cite{Goan16:032113} 
The same holds for the solutions of the respective equations, the memory effects present in the 
non-Markovian description represent higher order terms in the system-leads coupling. 

From our modeling efforts it is clear that the character of the transport into and through the system
is highly tunable with, or strongly dependent on, the plunger gate voltage, the photon 
frequency, and the initial number of photons present in the cavity. This fact underlines the importance of a 
computationally effective approach to explore the dynamic properties of the system,\cite{2016arXiv161003223J} 
and the possibility to enhance further the description of the details of the underlying physical processes 
beyond the presently implemented sequential tunneling to high order.\cite{Goan16:032113}

With the large number of tunable parameters available in the system and our emphasis on 
geometrically dependent effects, a bit neglected property of quantum optical transport systems,   
we observe that in general the properties of the system have to be explored further  
before making statements about possible technical applications and devices at this moment.
We notice a strongly growing interest in implementing quantum optical systems
and quantum computing in solid-state systems spurring researchers to model complex 
interacting systems.

\begin{acknowledgments}
This work was financially supported by the Research Fund of the University of Iceland,
the Icelandic Research Fund, grant no.\ 163082-051, 
and the Icelandic Instruments Fund. We also acknowledge support from the computational 
facilities of the Nordic High Performance Computing (NHPC), and the Nordic network
NANOCONTROL, project no.: P-13053. HSG acknowledges support from MOST, Taiwan, under grant no.\
103-2112-M-002-003-MY3.
\end{acknowledgments}

%
%----------------------------------------------------------------------------------------
%
\bibliographystyle{apsrev4-1}
%\bibliography{mod_qd}
%

%
%
%----------------------------------------------------------------------------------------
%
\end{document}